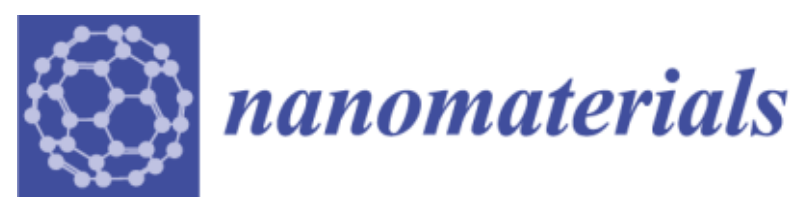

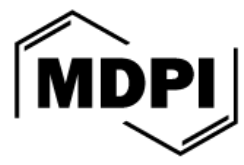

*Article*

# High-Field EPR/ENDOR of N/Be Centers for Defect Engineering in 6*H*-SiC

**Yuliya Ermakova [1,*], Ekaterina Dmitrieva [1], Margarita Sadovnikova [1], Fadis Murzakhanov [1,*], George Mamin [1], Sergey Nagalyuk [2], Evgeny Mokhov [2] and Marat Gafurov [1,3]**

[1] Institute of Physics, Kazan Federal University, Kremlyovskaya 18, Kazan 420008, Russia; dev600@mail.ru (E.D.); margaritaasadov@gmail.com (M.S.); george.mamin@kpfu.ru (G.M.); marat.gafurov@kpfu.ru (M.G.)

[2] Ioffe Institute, Polytekhnicheskaya, 26, St. Petersburg 194021, Russia; snagalyuk@gmail.com (S.N.); mokhov@mail.ioffe.ru (E.M.)

[3] Higher School of Cybertechnology, Mathematics and Statistics, Plekhanov Russian University of Economics, Stremyannyy Pereulok, 36, Moscow 115093, Russia

* Correspondence: yliyaermakova@gym5cheb.ru (Y.E.); murzakhanov.fadis@yandex.ru (F.M.)

**Abstract**

Silicon carbide (SiC) in its various structural modifications is widely used in power semiconductor electronics, operating under extreme conditions of high temperature, high voltage, and intense radiation. The discovery of spin defects ($S > 0$) with unique optical and coherent properties has further positioned SiC as a promising platform for quantum technologies. Here, we investigate a 6*H*-SiC single crystal co-doped with nitrogen and beryllium at concentrations of $10^{18}$ cm$^{-3}$, using continuous-wave and pulsed electron paramagnetic resonance (EPR) and electron–nuclear double resonance (ENDOR). To enhance spectral resolution, experiments were conducted in the W-band (94 GHz; $B = 3.4$ T). Pulsed EPR identified nitrogen donors and beryllium acceptors in various lattice positions, allowing for the determination of their phase coherence and spin–lattice relaxation times. ENDOR measurements elucidated the electron–nuclear interactions with the local silicon and carbon environment, including distant coordination spheres. The observed hyperfine structures indicated highly delocalized spin density within the supercell. The TRIPLE resonance spectra verify coupled nuclear spin subspaces from different coordination spheres due to defect spin density. These results demonstrate the feasibility of incorporating dual impurities with distinct functional roles while preserving the crystal lattice's structural features.

## 1. Introduction

Silicon carbide (SiC) of the 6*H* polytype is a wide-bandgap semiconductor ($E_g = 3$ eV) [1] that has attracted attention for power electronics and extreme-environment applications owing to its unique physical properties, such as high thermal conductivity, mechanical strength, and radiation hardness [2]. Over the past decade, interest in this material has expanded significantly with the development of quantum technologies, since 6*H*-SiC can serve as a host matrix for point-defect centers with controllable spin states at room temperature [3].

Nitrogen is the primary donor impurity in 6*H*-SiC, providing n-type conductivity by creating shallow levels near the conduction band [4–6]. Substituting for structural carbon or silicon atoms, nitrogen forms donor centers, and in combination with silicon vacancies, it creates negatively charged high-spin nitrogen vacancy (NV) centers with axial symmetry. NV centers are highly promising for quantum applications since their electron spin can be optically initialized and read out by ODMR spectroscopy [7,8]. The spin–lattice and spin–spin relaxation times in isotopically enriched crystals reach milliseconds and tens of microseconds, respectively. Meanwhile, beryllium is an acceptor impurity used to achieve p-type conductivity [9] and to act as a tool for controlling electronic properties and creating compensated structures that are necessary for semi-insulating substrates and fine-tuning the energy spectrum. In 6*H*-SiC crystals, beryllium can occupy various inequivalent lattice sites, forming both shallow and deep acceptor centers; their structural and spin properties depend significantly on temperature and the doping method [10]. Thus, nitrogen and beryllium in 6*H*-SiC perform complementary functions: the former serves as a key constituent for quantum defects with long-lived spin memory [11], while the latter enables the formation of p–n junctions [12,13] and control of conductivity type. This synergy paves the way for the integration of quantum elements with semiconductor electronics (see Figure 1a). A comprehensive understanding of the physical nature, local structure, and relaxation characteristics of these impurities is fundamental to the design of silicon carbide-based quantum devices.

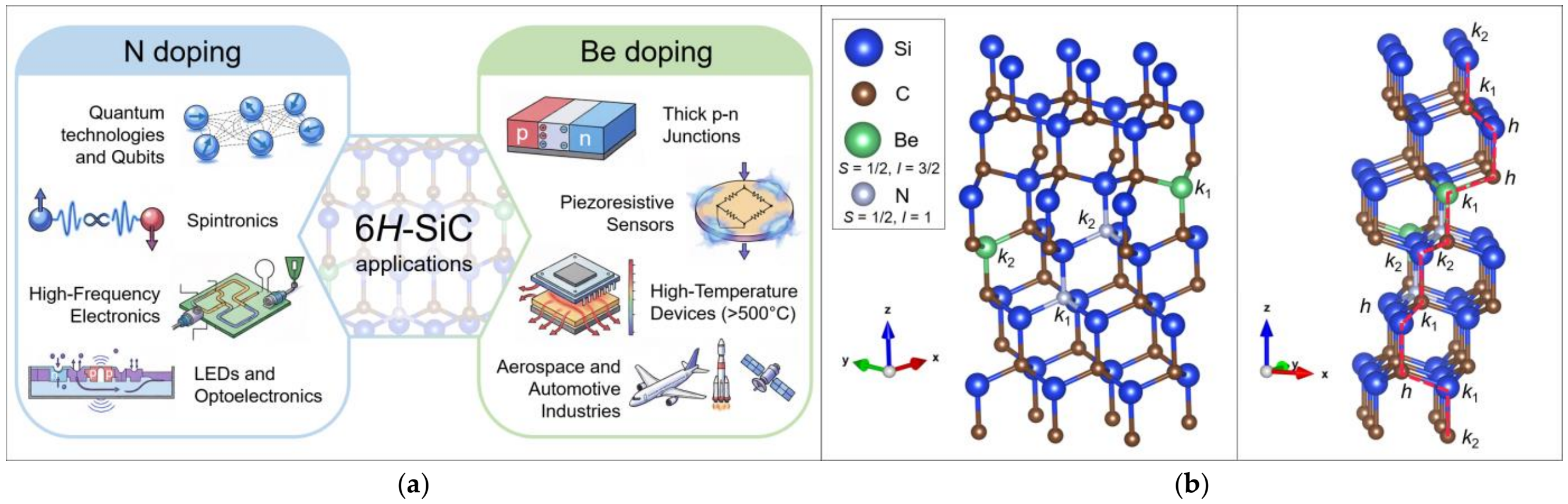

(**a**) (**b**)

**Figure 1.** (**a**) Applications of 6*H*-SiC doped with beryllium and nitrogen. (**b**) Crystal structure of Be- and N-doped 6*H*-SiC: (**left**)—Be and N impurities in the $k_1$ and $k_2$ lattice configurations; (**right**)—the same structure in an alternative orientation, where dashed lines indicate the designation of configurations depending on the impurity positions.

Optically active point defects with addressable spin states are essential for quantum technologies [14,15]. They show great potential in communication networks [16,17], computing [18], and high-resolution sensing [19]. Notably, many of these systems operate at room temperature [20]. Spin qubits are highly attractive due to their long coherence times and optical controllability. Furthermore, they are compatible with established semiconductor fabrication processes. This makes them strong candidates for scalable quantum platforms [21–23].

The nitrogen vacancy center in diamond remains the most extensively studied system. It has reached considerable technological maturity over the past decade [24]. However, there is a growing demand for alternative materials. This has motivated interest in silicon carbide. It is a technologically mature wide-bandgap semiconductor that hosts a rich family of analogous lattice defects [25]. SiC combines well-established wafer growth techniques with the ability to accommodate numerous spin-active centers, positioning it as a bridge between conventional electronics and quantum applications.

High-energy particle irradiation efficiently generates primary point defects uniformly throughout the SiC bulk [26]. These include silicon vacancies ($V_{Si}$; $S$ = 3/2) [27] and carbon vacancies ($V_C$) [28], which subsequently migrate and combine during thermal annealing to form paired spin complexes such as divacancies (VV; $S$ = 1) [29]. This approach offers high uniformity and reproducibility, enabling controlled fabrication of dense defect ensembles. Silicon vacancy represents the most thoroughly characterized defect and has been validated as a functional qubit [30]. Using synchronous detection, nanotesla-scale magnetic field sensitivity has been demonstrated for silicon vacancies in silicon carbide [31].

Electron paramagnetic resonance (EPR) and electron–nuclear double resonance (ENDOR) spectroscopy are powerful tools for probing electronic and nuclear spin subsystems. They allow the measurement of relaxation times and the extraction of spin-Hamiltonian parameters relevant to photoinduced processes. Pulsed ENDOR sequences further resolve electron–nuclear couplings with both the immediate atomic environment and distant coordination shells. This approach provides quantitative access to hyperfine and quadrupole interaction tensors. Combined with density functional theory (DFT) calculations, ENDOR has revealed the microscopic structure of axially symmetric NV centers at the *kk* and *hh* sites in 4*H*-SiC [32] and at the $k_2k_1$ and *hh* sites in 6*H*-SiC [33]. Ultimately, this structural insight enables the precise mapping of defect spin density within the crystal lattice, as demonstrated for silicon vacancies in 6*H*-SiC [34].

There are over 250 known polytypes of silicon carbide [35], differentiated primarily by how Si–C bilayers are stacked along the crystallographic c-axis. Variations in local symmetry and specific periodicity create three structurally distinct lattice positions for both silicon and carbon atoms. As shown in Figure 1b for the 6H polytype, these three unique positions are designated as *h*, $k_1$, and $k_2$, representing one hexagonal and two quasicubic sites, respectively. Irrespective of the atomic layout in a certain sublattice, the $k_1$ position is invariably situated next to an *h* position. Such labeling makes it easier to categorize the six potential configurations of nitrogen vacancy (NV) centers. For instance, the $k_2k_1$ designation describes a setup where a nitrogen atom sits at the $k_2$ site (within the carbon sublattice), while a silicon vacancy resides at the adjacent $k_1$ site (within the silicon sublattice).

Although vacancy-related defects in diamond and silicon carbide ~~(SiC)~~ have been widely investigated, the exact microscopic spin density distribution and the precise contribution of the silicon vacancy to the local electric field gradient (EFG) at inequivalent 6*H*-SiC sites remain poorly quantified. Furthermore, the literature currently lacks a direct comparison of the nuclear quadrupole interactions between isolated substitutional nitrogen and beryllium complexes situated in identical crystallographic environments. Performing this comparative analysis is crucial because it allows researchers to decouple the intrinsic host-matrix effects from the charge redistribution caused by the defect itself. Understanding this distinction is vital, as it directly dictates the feasibility of multi-level nuclear spin registers, the fidelity of optical initialization, and overall spin coherence. To address these gaps, we employ a combination of X- and W-band pulsed ENDOR spectroscopy in this study. This dual-frequency approach allows us to investigate electron–nuclear couplings in both weak and strong regimes, thereby facilitating the clear isolation of quadrupolar, dipolar, and isotropic interactions. Specifically, we analyze the EPR and ENDOR spectra of beryllium acceptors and substitutional nitrogen donors in 6*H*-SiC to extract quantitative values for their quadrupole and hyperfine interaction parameters. Ultimately, our primary goal is to determine the microscopic architecture of the NV defect. Additionally, by comparing the quadrupole coupling constant of the nitrogen donor, we aim to elucidate how the presence of the silicon vacancy modulates the local electric field gradient.

## 2. Materials and Methods

### *2.1. Sample*

Beryllium was introduced into 6*H*-SiC samples by vapor diffusion. The impurity source was high-purity metallic Be. Diffusion was conducted in closed tantalum carbide containers at 2000 °C for 1 h. The initial SiC samples had n-type conductivity with an uncompensated donor concentration ($N_d$-$N_a$) of 5 × $10^{18}$ $cm^{-3}$. After diffusion, the samples acquired p-type conductivity.

The synthesized material is shown in Figure 2a. After diffusion, the material darkened significantly, indicating a high Be impurity concentration. According to Hall measurements, the acceptor Be concentration was 10 $cm^{-3}$. However, since Be is an amphoteric impurity in SiC, a significant fraction occupies interstitial sites and acts as a deep donor.

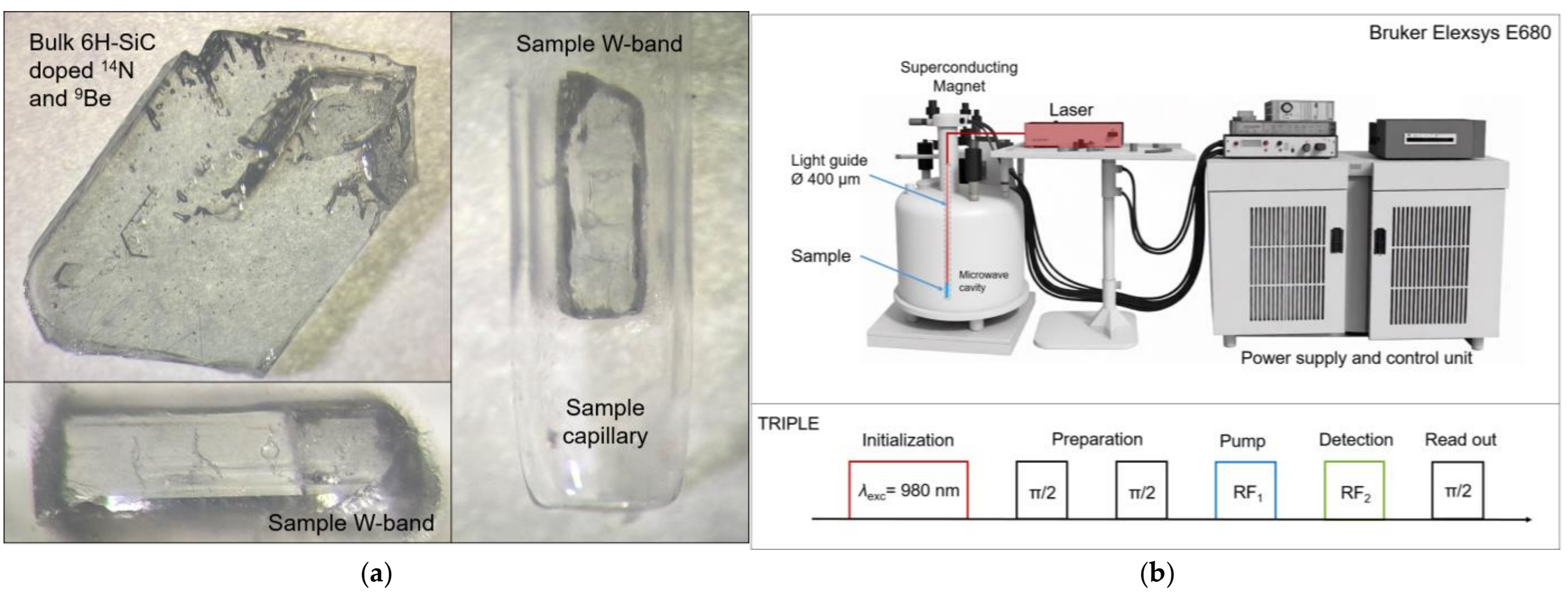

(**a**) (**b**)

**Figure 2.** (**a**) Photographs of the 6*H*-SiC sample. (**b**) Upper panel—W-band EPR setup used in the experiments; lower panel—pulse sequences employed for the measurements.

### *2.2. Experimental Approaches*

The magnetic resonance measurements were performed using a Bruker Elexsys E580/E680 spectrometer (Bruker, Karlsruhe, Germany) operating at microwave frequencies of ν = 9.6 GHz (X-band) and 94 GHz (W-band) (see Figure 2b). The EPR data were collected in pulsed mode via electron spin echo (ESE) detection as a function of magnetic field, using the Hahn pulse sequence $\pi_{MW/2}$–τ—$\pi_{MW}$—τ—ESE, with pulse durations of $\pi_{MW/2}$ = 40 ns and π = 1500 ns. The Mims pulse sequence ($\pi_{MW/2}$–τ—$\pi_{MW/2}$–$\pi_{RF}$—$\pi_{MW/2}$–τ—ESE) was used to record ENDOR spectra, with a radiofrequency (RF) pulse duration of $\pi_{RF}$ = 72 µs. An additional RF source (1–250 MHz) with an output power of 150 W was used to initiate nuclear magnetic resonance (NMR) transitions. Low-temperature experiments were conducted using a helium-flow cryostat from Oxford Instruments (Abingdon, Oxfordshire, UK). Optical excitation of the samples was implemented by the solid-state diode-pumped laser (*λ* = 980 nm) with an output power of up to 500 mW. The pulse programmer allowed the effective laser power to be reduced to a nominal 50 mW. Based on the sample volume and the heat capacity of SiC, the estimated temperature rise during a 5 ms optical pulse was approximately 1.5 K. The high helium flow in the cryostat ensured efficient dissipation of optically induced heat, resulting in negligible deviations and an experimental error of less than 1% for ENDOR measurements.

## 3. Results

The continuous-wave (CW) X-band EPR spectrum of the sample at room temperature is shown in Figure 3a. A single line is observed in the g-factor region for an isolated center, accompanied by four weakly resolved, low-intensity lines. The EPR signal was obtained at a high microwave power, indicating short relaxation times and no saturation. This signal can be tentatively attributed to resonant transitions of beryllium centers involving the $^{9}$Be isotope, which has a nuclear spin $I = 3/2$. No additional microwave absorption signals were observed in the rest of the magnetic field range. Subsequently, to lengthen relaxation times and ensure an optimal signal-to-noise ratio, experiments were carried out at low temperatures. The pulsed X-band EPR spectrum of the nitrogen donor in 6*H*-SiC is shown in Figure 3b. The signal comprises three equidistant lines, with an isotropic g-factor of approximately 2.004 and characteristic hyperfine splitting. The spectral shape remains independent of the crystal rotation angle relative to the external magnetic field, as verified using a precision goniometer. The resonance line intensities showed no change upon optical irradiation across varying wavelengths and power levels.

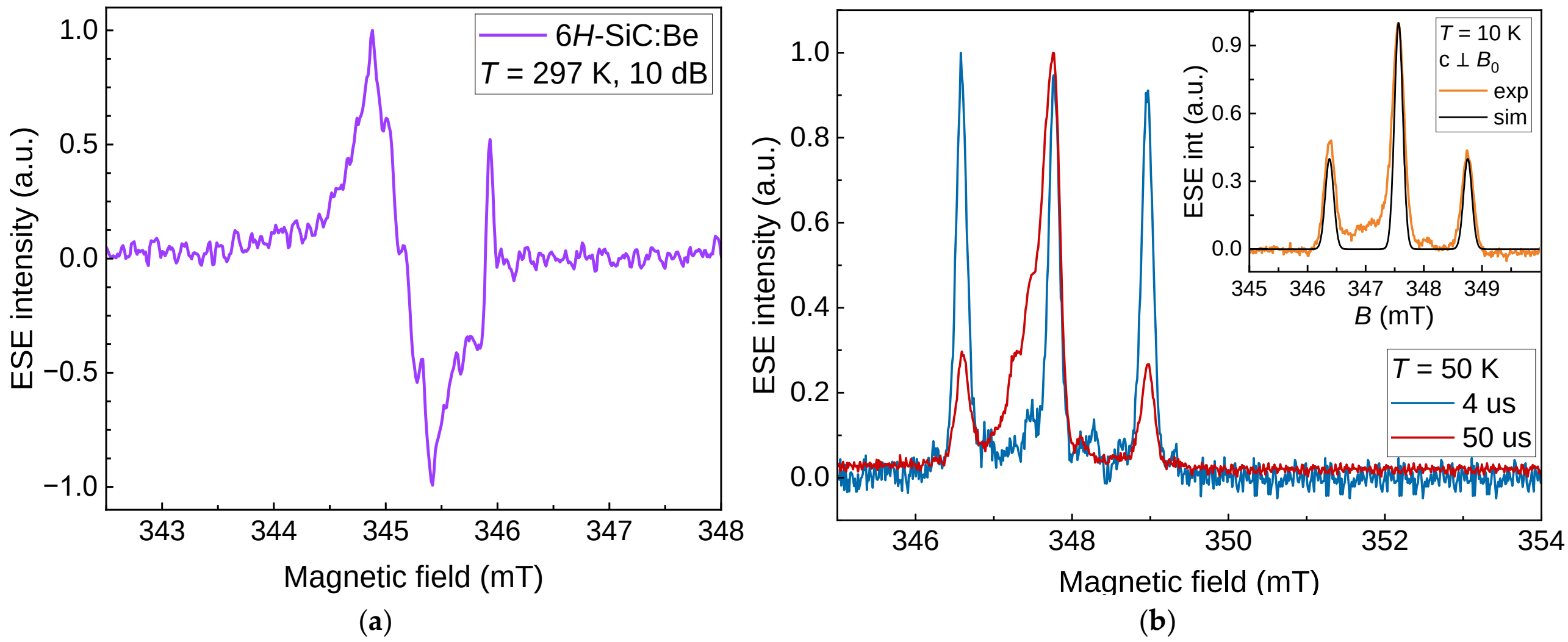

**Figure 3.** (**a**) Continuous-wave X-band EPR spectrum of beryllium centers at room temperature of the 6*H*-SiC sample. (**b**) Pulsed X-band EPR (integrated ESE intensity as a function of the magnetic field) for nitrogen donors at 50 K, with the inset showing the pulsed EPR spectrum at 10 K and the dashed lines indicating simulations based on the spin Hamiltonian values.

To describe the experimental results for point defects with electronic spin $S = ½$, the following axial spin Hamiltonian is employed:

$$H = g_{||}\mu_B \boldsymbol{B}\cdot\boldsymbol{S} + +A_{||}S_zI_z + A_{\perp}\left(S_xI_x + S_y\mathrm{I}_y\right), \quad (1)$$

where $g_{||}$ is the spectroscopic splitting factor for the orientation with the crystallographic *c*-axis parallel to the external magnetic field $B_0$ ($\vartheta = 0°$); $\mu_B$ is the Bohr magneton; $S_{x,y,z}$ and $I_{x,y,z}$ denote the Cartesian components of the electron and nuclear spin operators; and *A* represents the hyperfine interaction tensor.

The experiments were carried out with varying timing parameters of the pulse sequence, specifically by adjusting the repetition time (delay between pulse trains) and interpulse intervals (τ). Such variation in the spectrometer settings allowed us to selectively detect different paramagnetic centers with similar spectroscopic values (spin Hamiltonian values) but distinct relaxation dynamics (transverse and longitudinal relaxation times). Moreover, over a wide temperature range at X-band frequencies, beryllium centers were not detected in either CW or pulsed modes. The simulated EPR spectra of the nitrogen donors, calculated using spin Hamiltonian (1), are shown as dashed-dotted lines. EPR and ENDOR experiments were subsequently performed in the W-band to enhance sensitivity and spectral resolution. As shown in Figure 4a, the EPR spectrum of the nitrogen donor resolves into two distinct triplet patterns due to differences in their g-factors. The temperature dependence of the W-band pulsed EPR spectra is also presented. The EPR signals from beryllium impurities become detectable only upon significant cooling of the crystal temperature.

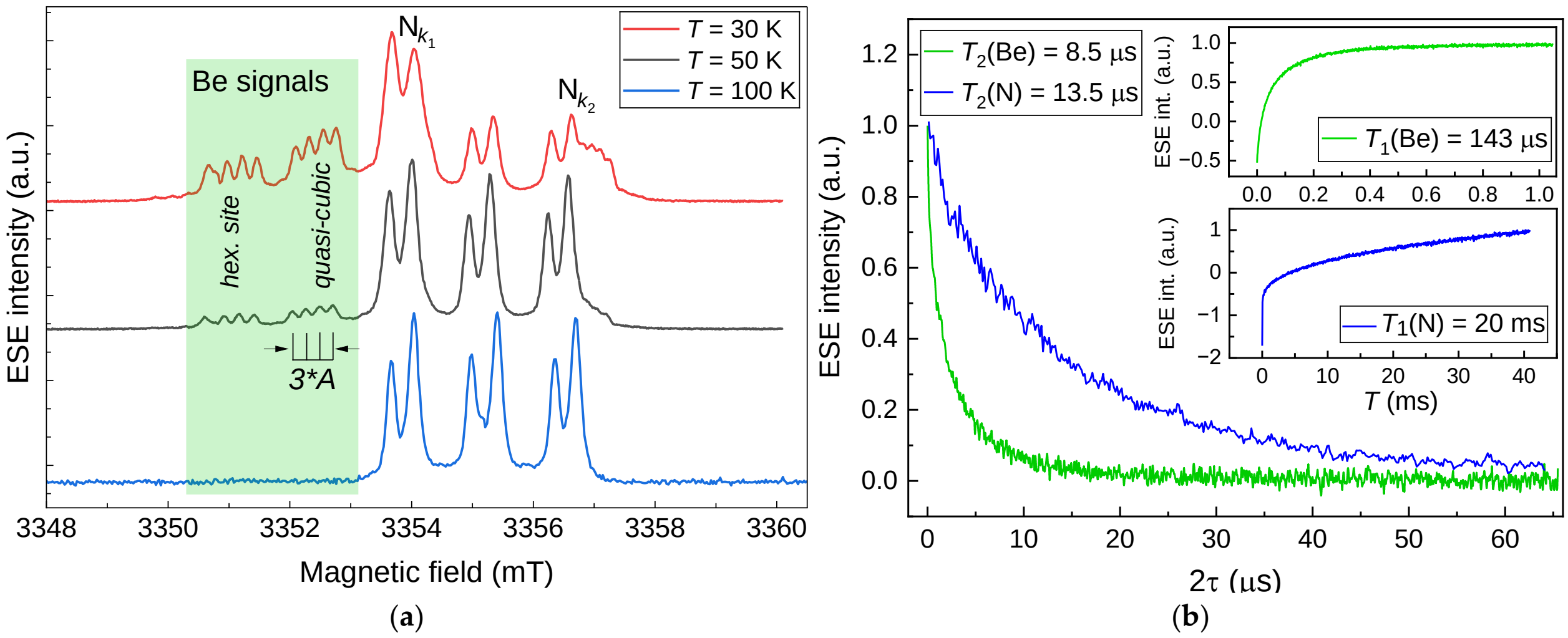

**Figure 4.** (**a**) Pulsed W-band EPR spectra for beryllium and nitrogen spin centers in 6*H*-SiC as a function of crystal temperature. The green (low-field) region highlights the beryllium acceptors. (**b**) Relaxation curves for nitrogen and beryllium centers. Main plot: decay of transverse magnetization due to spin–spin interaction. Inset: recovery of longitudinal spin magnetization characterized by the spin–lattice relaxation time.

The unambiguous assignment of the beryllium-related configurations to specific lattice sites is achieved by analyzing the high-field W-band (94 GHz) EPR spectra. While standard low-frequency X-band measurements exhibit severe spectral overlap due to minimal variations in the g-factors of distinct sites, the high Zeeman resolution at 3.4 T splits the resonance signals into two well-defined sets of four-line quartets determined by the $^{9}$Be nuclear spin ($I = 3/2$). The low-field quartet is characterized by a larger g-factor and a larger hyperfine splitting constant of 2.45 mT (686 kHz). This spectroscopic profile matches a stable, non-averaged axial configuration corresponding to the shallow beryllium acceptor at the hexagonal substitution site (the *h*-site), where the electron spin density remains localized along the crystal c-axis. Conversely, the high-field quartet exhibits a lower g-factor and a reduced hyperfine splitting constant of 2.2 mT (616 kHz). This reduction provides direct evidence of the dynamic Jahn–Teller effect inherent to the quasi-cubic substitution sites (the $k_1$ and $k_2$ positions). At the measurement temperature, rapid hopping of the beryllium ion between distorted local energy minima partially averages out the electronic wave function, scaling down the effective spin density directly at the $^{9}$Be nucleus. This distinct evolution of g-factors and hyperfine constants enables a highly confident differentiation between the shallow axial and quasi-cubic beryllium spin centers in the 6*H*-SiC matrix.

The phase coherence time, which characterizes the decay of the transverse magnetization due to spin–spin interactions, was determined at 30 K (see Figure 4b). The spin–spin relaxation rate was measured specifically with a dark signal to obtain the “pure” (intrinsic) coherence time of paramagnetic centers, where the magnetization decay is governed solely by intrinsic defect interactions within the crystal lattice. The transverse relaxation time curves follow a monoexponential function, indicating a single dominant spin–spin relaxation mechanism. Selective measurements employing microwave pulses with a narrow excitation bandwidth allowed for the determination of the relaxation times for nitrogen and beryllium centers, which differed by a factor of two (N: $T_2$ = 13.5 μs; Be: $T_2$ = 8.5 μs). The temperature dependence of the transverse relaxation time is expected to be influenced by spin–lattice relaxation (which shortens $T_2$), according to the relation $1/T_2^* = 1/T_2 + 1/T_1$.

The transverse relaxation time $T_2$ for nitrogen and beryllium impurities in 6*H*-SiC is determined by a complex interplay of physical and structural factors, each contributing to the dephasing of the spin system. As the temperature increases from 100 to 300 K, $T_2$ shortens significantly. Additionally, for beryllium centers, local atomic configurations undergo rearrangement at 10 and 50 K, further modulating the relaxation. Increasing the concentration of paramagnetic impurities enhances dipole–dipole interactions between spins. This effect is more pronounced for nitrogen centers at high doping levels, whereas for beryllium, it depends on whether shallow or deep acceptor states are involved. The most profound effect is exerted by the isotopic composition of the crystal; substituting $^{29}$Si nuclei (non-zero nuclear spin) with $^{28}$Si ($I = 0$) weakens the nuclear spin bath. For NV centers in isotopically enriched 6*H*-$^{28}$SiC, $T_{2^*}$ can be increased by several factors compared with the natural isotopic composition. The local nuclear environment contributes through spin diffusion; nitrogen occupying quasi-cubic or hexagonal sites yields different hyperfine structures and, therefore, different dephasing rates. Shallow $Be_{Si}$, with the majority of the spin density on the nearest carbon, behaves differently than a deep $Be_{Si}$-C vacancy complex, where the unpaired electron is distributed over three silicon atoms, altering the efficiency of the spin–spin interaction. The method of beryllium introduction (diffusion or doping during growth) also affects $T_2$, as it generates different concentrations of vacancies and associated defects, which serve as additional dephasing channels. Irradiation and subsequent annealing can both create new centers and alter local fields, which also impact the relaxation times. The observed $T_2$ represents the combined result of all the aforementioned mechanisms, with their relative importance varying significantly depending on the sample composition and experimental conditions. Therefore, the practical use of 6*H*-SiC in quantum sensors or qubits requires a tailored approach, including temperature selection, doping control, isotopic purification, and minimization of structural defects. This enables the targeted extension of the coherence times for both nitrogen donors and beryllium acceptors.

The longitudinal relaxation time $T_1$ for nitrogen and beryllium impurities in 6*H*-SiC is determined by the interaction of the spin system with the lattice. For beryllium centers ($T_1$ = 143 μs), $T_1$ is significantly shorter than for nitrogen donors ($T_1$ = 20 ms) under identical conditions (see inset in Figure 4b). This difference arises from the nature of the unpaired electron wave functions. For nitrogen, which substitutes for carbon or silicon, the electron density is widely distributed, resulting in relatively weak coupling to phonons. For beryllium, in contrast, the electron is strongly localized on nearby atoms, especially in deep acceptor complexes, which strengthens the spin–lattice interaction. As the temperature increases, $T_1$ naturally decreases due to an increase in the population of phonon modes. For beryllium, this decrease begins at lower temperatures due to the presence of low-lying excited states. Impurity concentration also contributes through cross-relaxation and spin diffusion; for nitrogen, this mechanism becomes noticeable only at high doping levels, whereas for beryllium, it can operate even at moderate concentrations due to the specific nature of exchange interactions. The nuclear spin bath, determined by the isotopic composition of the crystal, affects $T_1$ to a lesser extent than $T_2$. Hyperfine field fluctuations can still provide an additional relaxation channel, although for beryllium, this contribution is often overshadowed by stronger phonon processes. For nitrogen centers occupying different positions in the hexagonal lattice, $T_1$ varies only slightly. In contrast, the transition from shallow $Be_{Si}$ to the deep $Be_{Si}$-$V_C$ complex is accompanied by a sharp drop in $T_1$ due to a change in local symmetry and increased spin–orbit mixing. Defects arising during doping or irradiation create local deformations and additional levels in the bandgap. This is especially critical for beryllium, as its acceptor states are extremely sensitive to stress and can open additional coherence loss channels. The crystal growth method and heat treatment affect the degree of compensation and the distribution of impurities, which, in turn, influences the lifetimes of excited states and the probability of multiphonon transitions. The phonon mechanism involving local vibrational modes remains dominant for beryllium, leading to a consistently shorter $T_1$ compared with nitrogen, where weaker interactions with acoustic phonons predominate. These differences must be taken into account when selecting the operating parameters of 6*H*-SiC-based devices, such as when designing quantum sensors or memory elements for which maximum relaxation times are required.

The experimental coherence times obtained at a relatively high operating temperature of 30 K, specifically $T_2$ = 13.5 μs for the nitrogen donors and $T_2$ = 8.5 μs for the beryllium acceptors, demonstrate a strong potential for solid-state quantum technology implementations. For wide-bandgap semiconductor acceptor states, which are typically subject to fast spin–orbit and phonon-induced relaxation pathways, a coherence window of 8.5 μs at 30 K indicates a remarkably robust spin system. In quantum memory architectures, these electronic coherence times are sufficient to serve as a fast interface for quantum state initialization and microwave readout sequences. The coherent spin state can then be mapped via electron–nuclear transitions onto the nuclear spin of the $^9$Be nucleus ($I$ = 3/2) or surrounding lattice nuclei. Because nuclear spins are effectively decoupled from environmental electric field fluctuations, their coherence lifetimes are several orders of magnitude longer than electronic ones, which ensures viable long-term quantum storage at 30 K. In quantum-sensing applications, the strong spin–lattice coupling governed by the Jahn–Teller distortion of the beryllium center enables high sensitivity to external strain, acoustic fields, and local electric gradients. The measured microseconds-long coherence times at 30 K guarantee that coherent sensing protocol sequences can be fully executed well within the dephasing limit. Furthermore, the realization of these spin properties at 30 K significantly reduces cryogenic cooling constraints compared with conventional solid-state qubits that demand mK refrigeration. This operating regime allows the deployment of compact closed-cycle cryocoolers, facilitating the integration of these co-doped spin networks into hybrid semiconductor systems using mature SiC microelectronic fabrication platforms.

Figure 5a displays the radiofrequency ENDOR spectrum of the nitrogen donor, which was acquired at a constant magnetic field corresponding to the high-field EPR transition, utilizing a 25 MHz microwave excitation bandwidth. The presence of two distinct line pairs indicates that the nitrogen donor atoms reside in two different quasicubic lattice positions, specifically the $k_1$ and $k_2$ sites, having substituted for carbon atoms during crystal growth. As reported in [36], the $N_{k1}$ site is associated with a shallower donor level at $E_1$ = 137.6 meV, whereas the $N_{k2}$ site corresponds to a deeper donor level at $E_2$ = 142.4 meV. Despite minor variations in local symmetry and environment, the overall magnitude of the hyperfine interaction remains nearly identical for both sites: $A(N_{k1})$ = 33.23(3) MHz and $A(N_{k2})$ = 33.55(3) MHz (Table 1). In standard EPR spectra, this minute difference of merely a few hundred kilohertz leads to inhomogeneous line broadening, thereby obscuring the resolution of these individual crystallographic sites. Furthermore, fluctuations in the intensity of the central EPR line could be attributed to an overlapping signal from the hexagonal $N_h$ site, which possesses a significantly smaller hyperfine constant of $A \approx 2.5(1)$ MHz. Conversely, ENDOR spectroscopy successfully resolves these subtle variations in electron–nuclear coupling, benefiting from the inherently narrow linewidths of nuclear transitions and the application of extended RF pulses (72 μs). At the specific fixed magnetic field of 340 mT, the hyperfine coupling strength is roughly an order of magnitude greater than the $^{14}$N Larmor frequency ($\nu_L = 1$ MHz). Consequently, the RF transition lines are centered near $A/2 = 16.5$ MHz, with the separation between these lines matching twice the Larmor frequency ($2\nu_L = 2$ MHz). Such a ratio of interaction strengths is a hallmark of the strong coupling regime [37]. To accurately model the system, simulations of both the EPR and ENDOR spectra were conducted, incorporating all pertinent spin interactions. Although $^{14}$N possesses a nuclear spin of $I = 1$ and electric field gradients are present, no quadrupole splitting could be resolved. This indicates that the corresponding quadrupole coupling constant is likely smaller than 10 kHz, rendering it indistinguishable within the natural spectral linewidth.

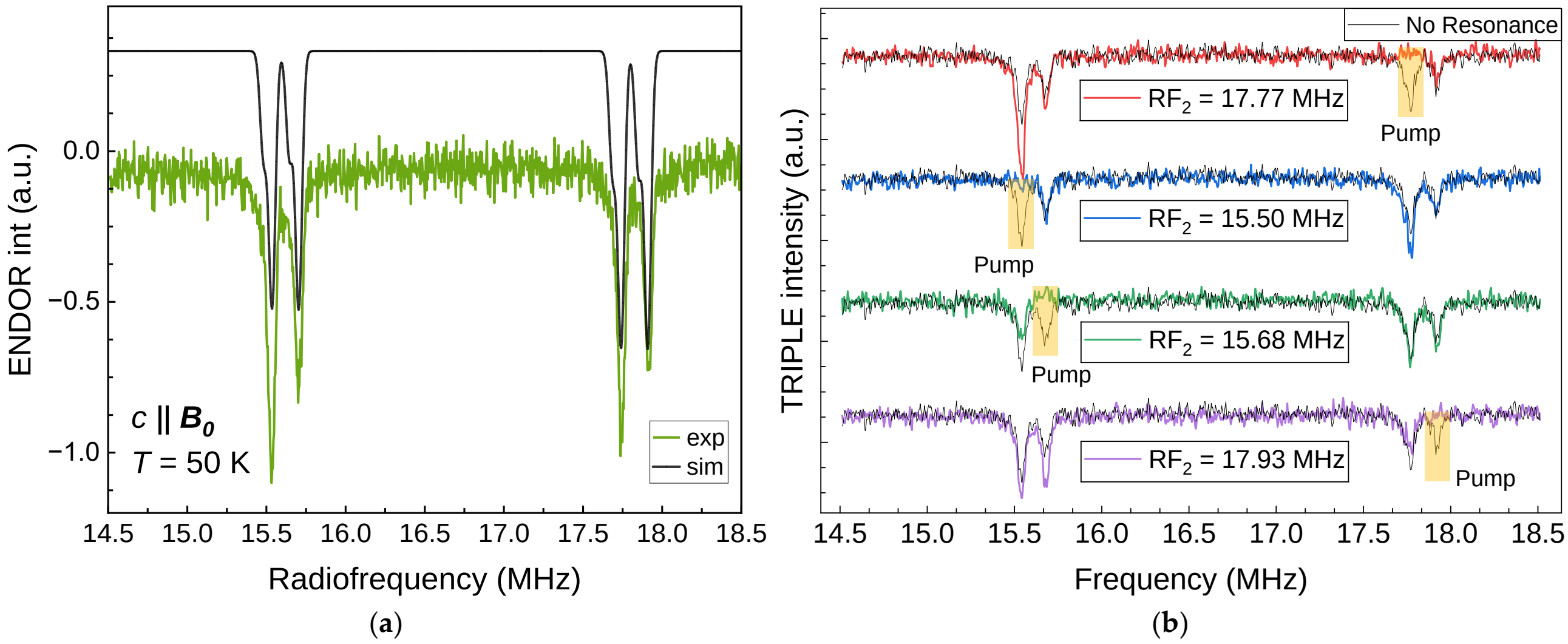

**Figure 5.** (**a**) X-band ENDOR spectrum at 50 K for the $^{14}$N isotope. The dashed line shows the simulation. (**b**) The results of selective spin excitation by an additional radiofrequency pulse (triple resonance) for specific NMR transitions are shown. The TRIPLE resonance spectra as a function of the fixed pump pulse are presented.

**Table 1.** Experimentally established values of electron–nuclear interactions for different types of defects in 6*H*-SiC ($N_{k1}$, $N_{k2}$, $NV_{k1k2}$) based on EPR and ENDOR spectroscopy analyses. The table defines the parameters of isotropic Fermi interaction ($a$), dipole–dipole coupling ($b$), and nuclear quadrupole splitting ($C_q$) for N (nitrogen donor defects) in configurations $k_1$ and $k_2$ ($N_{k1}$: $a = 33.23(3)$ MHz, $b \leq 10$ kHz, and $C_q \leq 10$ kHz; $N_{k2}$: $a = 33.55(3)$ MHz, $b \leq 10$ kHz, and $C_q \leq 10$ kHz). For comparison, the values of the parameter NV (nitrogen vacancy defect) in configuration $k_1k_2$ are given ($NV_{k1k2}$: $a = -1.22(1)$ MHz, $b \leq 15$ kHz, and $C_q = 2.426(3)$ MHz).

| Type of Defect | $a$ (MHz) | $b$ (kHz) | $C_q$ |
|---|---|---|---|
| $N_{k1}$ | 33.23(3) | ≤10 | ≤10 kHz |
| $N_{k2}$ | 33.55(3) | ≤10 | ≤10 kHz |
| $NV_{k1k2}$ | −1.22(1) | ≤15 | 2.426(3) MHz |

Even at the elevated magnetic field of 3.4 T, where the $^{14}$N Larmor frequency rises to roughly 10 MHz, the spin system continues to operate within the strong coupling regime. Under these conditions, the spectrum remains centered at 16.5 MHz but displays a substantially increased line splitting of 20 MHz. As depicted in the X-band ENDOR spectra (Figure 5a), only a single pair of resonance lines is detected. This observation is a direct consequence of the selective microwave excitation targeting nitrogen donors specifically at the $k_2$ lattice sites. Consequently, the investigated system can be accurately described as an electron spin ($S = 1/2$) coupled to a $^{14}$N isotope spin ($I = 1$). Given that the nitrogen donor occupies a substitutional carbon site, its interaction with the adjacent silicon atoms in the nearest-neighbor environment becomes a focal point of interest. Figure 5b shows TRIPLE resonance spectra obtained using an additional RF pulse in the Mims sequence to drive nuclear transitions between adjacent levels. For a given crystallographic site, an intensity redistribution occurs within a single nuclear multiplet. The small intensity transfer observed between adjacent multiplets indicates dipole–dipole coupling between nuclear spins located at different crystallographic sites of the nitrogen donors [4]. These TRIPLE resonance experiments confirm that the signals in the ENDOR spectrum originate from two inequivalent nitrogen donor centers, rather than from the quadrupole splitting of a single spin defect.

Figure 6a,b display the ENDOR spectra acquired at a fixed EPR transition of the nitrogen center, specifically targeting the Larmor frequency regions of $^{13}$C ($\nu_L$ = 35.8 MHz) and $^{29}$Si ($\nu_L$ = 28.3 MHz). Remarkably, despite the low natural isotopic abundances of $^{13}$C (1.1%, $I = 1/2$) and $^{29}$Si (4.8%, $I = 1/2$), their radiofrequency absorption signals remain highly pronounced. Unlike the $^{14}$N ENDOR spectra, the resonant nuclear transitions for $^{29}$Si feature a substantially greater number of components, characterized by a broad distribution of intensities and hyperfine coupling values. Specifically, ten distinct spectral features were resolved, with hyperfine coupling constants ranging from $A(1)$ = 140 kHz to $A(9)$ = 2.79 MHz, all demonstrating relatively narrow linewidths of roughly 5–10 kHz. Earlier X-band investigations utilized the Davies pulse sequence to map $^{29}$Si ENDOR spectra across various coordination shells under intermediate electron–nuclear coupling conditions, a regime where the Larmor frequency is comparable to the hyperfine coupling constant [4]. Conversely, the present work obtained high-frequency ENDOR spectra in the weak coupling regime, facilitated by a high magnetic field of 3.4 T and the application of the Mims pulse sequence. The Mims sequence is particularly advantageous for probing nuclei with small gyromagnetic ratios, such as $^{14}$N. Moreover, compared with the Davies method, this non-selective stimulated spin echo detection scheme provides superior sensitivity to the weak, closely spaced hyperfine interactions generated by distant nuclei.

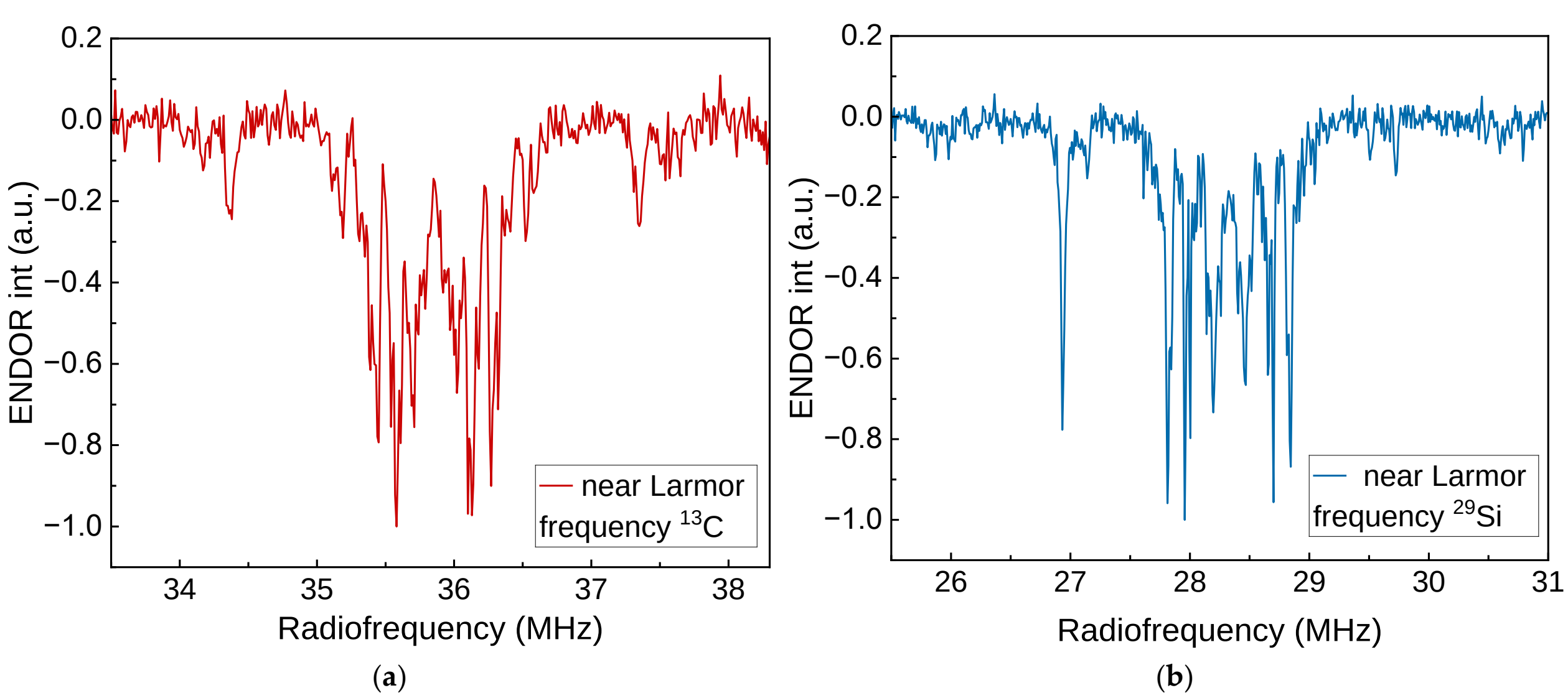

**Figure 6.** W-band ENDOR spectra at a fixed magnetic field for nitrogen donors in the region of the Larmor frequencies of the $^{29}$Si isotope with a nuclear spin of $I$ = ½ (**a**) and the $^{13}$C isotope with $I$ = ½ (**b**). The observed splittings arise from hyperfine interactions with different coordination shells due to the delocalized spin density of nitrogen donors within the crystal.

The wide variety of observed spectral features strongly indicates that the spin density of the nitrogen donor is highly delocalized throughout the unit cell. Indeed, estimations of the spatial extent of this defect reveal that the spin density spreads to at least twice the Bohr radius and beyond [38]. Rather than decaying monotonically with increasing distance $r$ from the substitutional site, the donor wave function exhibits a complex spatial profile. Furthermore, the electron spin density demonstrates pronounced anisotropy within the specific coordination shells encircling the nitrogen donor, a factor that directly modulates the hyperfine coupling constants. Such localized asymmetries, even at a constant electron–nuclear separation, can generate multiple inequivalent nuclear positions. This proliferation of distinct sites not only multiplies the observed spectral features but also adds considerable complexity to the interpretation of the ENDOR data. Ultimately, the hyperfine coupling constants identified in this study correspond to magnetic nuclei situated within a radial distance of up to 5 Å from the donor center. Previous studies have shown that the electron spin density of the nitrogen donor is distributed predominantly over silicon atoms in 4*H*-SiC and over carbon atoms in 6*H*-SiC [38,39].

In contrast with substitutional nitrogen donors, the spin Hamiltonian parameters associated with the nitrogen vacancy (NV) center demonstrate a pronounced angular dependence. As established in prior research [40], the zero-field splitting parameters were precisely quantified by acquiring EPR spectra at both canonical and intermediate crystal orientations relative to the external magnetic field, $B_0$. Furthermore, the alignment of the crystallographic c-axis within the $B_0$ reference frame plays a critical role in dictating the relaxation dynamics of the spin center, specifically its longitudinal ($T_1$) and transverse ($T_2$) relaxation times. To accurately model the electron–nuclear interaction (Table 1), three primary contributions were incorporated: the anisotropic dipolar coupling ($b$), the isotropic Fermi contact interaction ($a$), and the quadrupole coupling constant ~~($C_q$)~~. Here, the quadrupole frequency $Q$ is defined as $Q = (3eQ_NV_{zz})/(4I(2I-1)) = 3/(4I(2I-1))C_q$, with $Q_N$ representing the nuclear quadrupole moment. Specifically, the separation between RF absorption lines 1 and 3 corresponds to $2Q = 3.795$ MHz, which translates to a quadrupole coupling constant of $C_q = 2.53$ MHz. This stands in stark contrast with the substitutional nitrogen donor, where the corresponding value was previously reported to be merely $C_q = 7–9$ kHz [39]. Given that quadrupole splitting originates from the local electric field gradient ~~(EFG)~~, the principal source of this interaction in the NV center is the adjacent lattice defect, namely, the silicon vacancy ($V_{Si}$). The presence of this vacancy triggers a substantial charge redistribution around the $^{14}N$ nucleus, thereby generating an axially symmetric EFG characterized by an asymmetry parameter $\eta = (Q_{xx} - Q_{yy})/Q_{zz} = 0$ (where $Q_{jj}$ denote the diagonal components of the EFG tensor). Similar to the well-known nitrogen vacancy center in diamond, the hyperfine coupling in this system is predominantly isotropic and carries a negative sign ($a = -1.22$ MHz). This phenomenon is driven by the unique distribution of electron spin density across the nearest-neighbor carbon atoms encircling the silicon vacancy. Through a spin polarization mechanism, this density polarizes the core electron shells at the nitrogen site, ultimately yielding an effective negative isotropic hyperfine coupling constant. Mirroring the behavior of the nitrogen donor (4–9 kHz), the anisotropic dipolar contribution here is exceptionally minor, measuring only 10–15 kHz, rendering it negligible relative to the ENDOR spectral linewidth. Finally, the spin density distribution of the NV center is highly delocalized; the residual spin density localized directly on the $^{14}N$ nucleus is estimated to be ≤0.2%, a finding that aligns perfectly with prior assessments.

Because the nuclear charge distribution is inherently non-spherical, the nuclear energy levels exhibit a distinct orientation dependence relative to the local electric field gradient ~~(EFG)~~. As a highly localized property, the EFG is governed primarily by the charge distribution in the immediate vicinity of the nucleus [41]. For light nuclei like nitrogen, the Townes–Dailey model is routinely utilized to elucidate the mechanisms underlying EFG formation [42]. Within this simplified theoretical framework, the EFG is postulated to originate from uncompensated valence p-orbitals centered on the nucleus [43]. Since each valence p-orbital represents a non-spherical electron density distribution (or population imbalance), it contributes to the overall field gradient in direct proportion to its electron occupancy. Given that the chemical bonding in the 6*H*-SiC lattice is predominantly covalent (~90%) with only a minor ionic component (~10%) [44], the pristine host matrix does not produce a substantial EFG at the defect site. As a result, quadrupole features in the spectra remain unresolved, obscured by inhomogeneously broadened lines. However, a striking contrast emerges when examining the NV center within the same crystal: its quadrupole coupling constant ~~($C_q$)~~ is two orders of magnitude larger than that of the substitutional nitrogen donor, despite both sharing the same nuclear charge and spin ($I = 1$). This profound disparity in $C_q$ is directly attributable to a significant alteration of the principal EFG tensor component $V_{zz}$, which is induced by the adjacent silicon vacancy ($V_{Si}$) neighboring the substitutional nitrogen atom. Furthermore, this $V_{Si}$-driven asymmetric redistribution of charge is precisely what dictates the observed axial symmetry of the spin Hamiltonian tensors.

Experimental ENDOR spectra of beryllium centers in 6*H*-SiC, recorded in the W-band (95 GHz), exhibit a resolved multiplet structure, with observed hyperfine interaction (HFI) values ranging from 5.5 to 5.7 MHz (see Figure 7a). This value determines the degree of localization of the electron spin density of the paramagnetic center and is consistent with the off-center acceptor model $Be_{Si}$ [9,10]. Additional splitting of the resonance lines into components with a characteristic interval of about 220 kHz is interpreted as the contribution of the nuclear quadrupole interaction (NQI) for the $^9$Be isotope ($I = 3/2$). Such a small value of the quadrupole splitting ($Q = 220$ kHz) indicates a moderate asymmetry of the local electric field at the defect location. A comparative analysis of the obtained HFI and NQI constants with known literature data for shallow beryllium acceptor states in hexagonal silicon carbide polytypes allows us to attribute the center under study to one of the quasicubic positions ($k_1$/$k_2$) or a shallow axial center [10]. The high spectral resolution of the W-band allowed us to reliably separate the isotropic and anisotropic contributions to the spin Hamiltonian, minimizing the effects of inhomogeneous line broadening characteristic of measurements in the standard X-band [45].

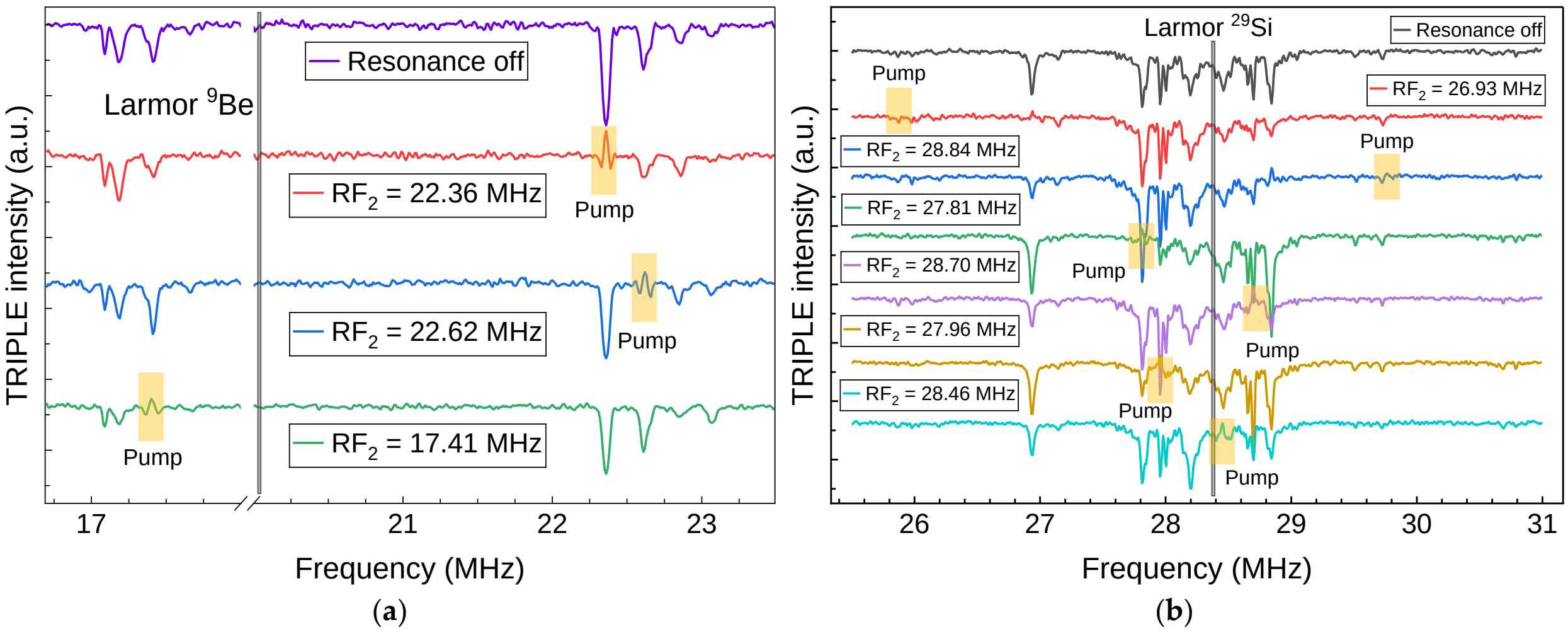

**Figure 7.** (**a**) W-band ENDOR spectra in the region of the Larmor frequency of the $^9$Be ($A$ = 5.5–5.7 MHz; $Q$ = 220 kHz) with $I$ = 3/2 and a 100% natural abundance. The results of excitation by an additional radiofrequency pulse (TRIPLE resonance) for specific NMR transitions are shown.

(**b**) TRIPLE resonance measurements of the RF resonant absorption signals in the region of the $^{29}Si$ Larmor frequency.

The microscopic interaction between the co-doped nitrogen and beryllium impurities within the 6*H*-SiC matrix is governed by structural relaxation, site selectivity, and charge compensation. Nitrogen acts as a shallow donor and preferentially substitutes for carbon atoms, causing a small, symmetric inward relaxation of the surrounding silicon coordination sphere due to its smaller atomic radius [38,46]. Beryllium functions as an acceptor and primarily occupies the silicon substitution sites or locates in interstitial spaces [47,48]. Owing to its small ionic radius, substitutional beryllium undergoes a pronounced off-center displacement along the crystal c-axis driven by a strong Jahn–Teller distortion, which stabilizes the defect into distinct shallow axial or quasi-cubic configurations. At the atomic and spin level, mutual charge compensation between the opposing donor and acceptor species plays a major role in shifting the Fermi level position. This electronic transfer directly controls the relative populations of the paramagnetic neutral donor and neutral acceptor states, which determines the intensity of the observable electron paramagnetic resonance profiles. Furthermore, the Coulombic attraction between the positively charged nitrogen donors and the negatively charged beryllium acceptors promotes the formation of spatially correlated donor–acceptor pairs. This close proximity modifies the local wave function overlap, establishing distinct spin–spin interactions and modulating the local hyperfine and quadrupole parameters resolved via high-field magnetic resonance techniques [49].

To probe the coordination shells surrounding the impurity centers, TRIPLE resonance experiments were additionally conducted (Figure 7b). The TRIPLE resonance results demonstrate intensity redistribution within a single spin multiplet. Selective radiofrequency excitation at a fixed RF frequency leads to a pronounced redistribution of intensities in the ENDOR spectrum. The observed behavior indicates an indirect coupling between inequivalent nuclei mediated by the delocalized spin density. The mechanisms underlying this population transfer between nuclear spin sublevels require further study. To validate a comprehensive model of the defect electronic structure in the 6*H*-SiC matrix, it is instructive to compare the spin Hamiltonian parameters of the studied beryllium centers with those of the classical nitrogen donors. While the investigated Be centers exhibit a hyperfine interaction of 5.5–5.7 MHz and a significant nuclear quadrupole splitting of approximately 220 kHz, the $^{14}N$ donor centers residing at the quasicubic lattice sites ($Nk_1$ and $Nk_2$) demonstrate a substantially larger HFI constant (about 33 MHz) accompanied by a strongly suppressed quadrupole splitting of only about 10 kHz. This pronounced divergence in both magnetic and electric interactions originates from the fundamentally different physical nature of the donor and acceptor ground states, as well as their respective local lattice relaxation mechanisms. The significantly higher HFI value of 33 MHz for the nitrogen donor is primarily driven by the Fermi contact interaction term, which is directly proportional to the unpaired electron spin density $|\Psi(0)|^2$ at the nucleus. Nitrogen substitutes for carbon ($N_C$) as a shallow donor, and its wave function retains a distinct s-like character with a high degree of localization on the central site. In contrast, the beryllium impurity acts as an acceptor, typically substituting for silicon ($Be_{Si}$). Although the ionic radius of $Be^{2+}$ (r = 0.27 Å for tetrahedral coordination [50,51]) is close to that of $Si^{4+}$ (r = 0.26 Å [50,51]), the ion undergoes a substantial off-center displacement from the regular silicon site. This off-center relaxation forces the unpaired hole density to localize predominantly on the dangling p-orbitals of the neighboring carbon atoms rather than on the beryllium nucleus itself, thereby strongly reducing the isotropic Fermi contact contribution to 5.5–5.7 MHz.

An inverse hierarchy is observed for the nuclear quadrupole splitting, which reflects the interaction of the nuclear electric quadrupole moment with the local electric field gradient tensor. For the $^{14}$N donor ($I$ = 1; $Q$ = 0.020 × $10^{-28}$ m$^2$), the near-vanishing quadrupole splitting (about 10 kHz) is a direct consequence of the highly symmetric, nearly tetrahedral distribution of the donor electron density and the minimal distortion of the surrounding silicon shell. In contrast, the $^9$Be acceptor nucleus ($I$ = 3/2; $Q$ = 0.053 × $10^{-28}$ m$^2$) experiences a drastically enhanced EFG. This enhancement is doubly driven: first, by the intrinsically larger nuclear quadrupole moment of the beryllium isotope, and second, by the severe axial or lower-symmetry crystalline distortion induced by the off-center position of the $Be_{Si}$ ion. The resulting strong local lattice relaxation destroys the local cubic symmetry, generating a substantial EFG at the nucleus that manifests as a well-resolved 220 kHz quadrupole splitting in the W-band ENDOR spectra. These findings unambiguously demonstrate that high-frequency ENDOR spectroscopy serves as an ultrasensitive probe of local atomic displacements and wave function symmetries in wide-bandgap semiconductor crystals.

The distinct contrast between the nitrogen and beryllium centers highlights the unique complementary nature of the 6*H*-SiC platform for next-generation technologies. The highly localized and symmetric nitrogen donors, with their stable spin states, represent excellent candidates for quantum information processing and quantum sensing, where protecting spin coherence from electric field fluctuations is crucial. Simultaneously, the high sensitivity of the beryllium acceptor to its local electrical environment provides a deep understanding of semiconductor doping and lattice relaxation at the atomic scale. Harnessing both types of defects unlocks the full potential of silicon carbide, paving the way for advanced hybrid quantum–semiconductor devices where robust quantum memories and precise electronic control are integrated into a single, high-performance crystal platform.

## 4. Conclusions

In this work, a comprehensive high-frequency EPR and ENDOR investigation of a 6*H*-SiC single crystal simultaneously co-doped with nitrogen and beryllium was performed, yielding the following key findings:

- The structurally non-equivalent positions of nitrogen and beryllium spin centers were established, revealing a fundamental dichotomy in their physical nature. The nitrogen donor exhibits a highly symmetric, localized wave function with long spin coherence times ($T_2$), making it ideal for robust quantum memory. In stark contrast, the beryllium acceptor undergoes significant off-center distortion, leading to a highly delocalized hole density and shorter-lived states, rendering it exquisitely sensitive to local lattice strain.
- High-frequency (W-band) ENDOR spectroscopy enabled the mapping of the local atomic environment with sub-nanometer resolution. The observed hyperfine structures reflect the distinct spin density distributions: a localized density for the N center and a highly distributed spin density over neighboring carbon coordination spheres for the Be center. This confirms the massive local electric field gradient generated by the structural distortion of the Be center.
- Triple-resonance (TRIPLE) spectroscopy revealed coherent internuclear dipole–dipole interactions mediated by the defect spin density. The successful mapping of interconnected nuclear spin subspaces from different coordination spheres demonstrates the high quantum control potential and allows for the creation of multi-level energy architectures for quantum manipulation.

Ultimately, the distinct contrast between the environmentally stable nitrogen donors and the highly sensitive beryllium acceptors highlights the unique, complementary nature of the dual-doped 6*H*-SiC platform. Harnessing both types of defects within a single crystal matrix paves the way for advanced hybrid quantum–semiconductor devices, seamlessly integrating robust quantum information processing with nanoscale quantum sensors. Combined with the advanced technological maturity of silicon carbide and the exceptional spectral resolution achieved in this work, 6*H*-SiC is established as a versatile and robust platform for next-generation scalable quantum technologies.

**Author Contributions:** Conceptualization, M.G. and E.M.; methodology, G.M. and M.S.; software, G.M. and Y.E.; validation, F.M. and M.G.; formal analysis, E.D. and G.M.; investigation, G.M., E.D. and M.S.; writing—original draft preparation, F.M. and G.M.; writing—review and editing, F.M., Y.E., M.G. and S.N.; project administration, F.M.; funding acquisition, F.M. All authors have read and agreed to the published version of this manuscript.

**Funding:** This research was funded by a subsidy allocated to Kazan Federal University for a state assignment in the sphere of scientific activities (project no. FZSM-2024-0010).

**Data Availability Statement:** The original contributions presented in this study are included in this article. Further inquiries can be directed to the corresponding author.

**Acknowledgments:** S. Nagalyuk (6*H*-$^{28}$SiC crystal synthesis) acknowledges support from the State Assignment for the Ioffe Institute (project no. FFUG-2024-0024, 'Functional Materials for Microelectronics and Photonics'). During the preparation of this manuscript, the authors used Alice AI for the purposes of refining and adjusting fragments of the graphical abstract and Figures 1a and 2b. The authors have reviewed and edited the output and take full responsibility for the content of this publication.

**Conflicts of Interest:** The authors declare no conflicts of interest.

## Abbreviations

The following abbreviations are used in this manuscript:

| | |
|---|---|
| EPR | Electron paramagnetic resonance |
| MW | Microwave |
| RF | Radiofrequency |
| NV | Nitrogen vacancy |
| ESE | Electron spin echo |
| ENDOR | Electron–nuclear double resonance |

## References

1. Capan, I. Electrically active defects in 3C, 4H, and 6H silicon carbide polytypes: A review. *Crystals* **2025**, *15*, 255.
2. Millan, J.; Godignon, P.; Perpina, X.; Perez-Tomas, A.; Rebollo, J. A survey of wide bandgap power semiconductor devices. *IEEE Trans. Power Electron.* **2014**, *29*, 2155–2163.
3. Lohrmann, A.; Johnson, B.C.; McCallum, A.J.; Castelletto, S. A review on single photon sources in silicon carbide. *Rep. Prog. Phys.* **2017**, *80*, 034502.
4. Savchenko, D.V.; Kalabukhova, E.N.; Pöppl, A.; Shanina, B.D. Electronic structure of the nitrogen donors in 6H SiC as studied by pulsed ENDOR and TRIPLE ENDOR spectroscopy. *Phys. Status Solidi B* **2012**, *249*, 2167–2178.
5. Savchenko, D.V.; Kalabukhova, E.N.; Kiselev, V.S.; Hoentsch, J.; Pöppl, A. Spin-coupling and hyperfine interaction of the nitrogen donors in 6H-SiC. *Phys. Status Solidi B* **2009**, *246*, 1908–1914.
6. Savchenko, D.V. The electron spin resonance study of heavily nitrogen doped 6H SiC crystals. *J. Appl. Phys.* **2015**, *117*, 4.
7. Du, J.; Shi, F.; Kong, X.; Jelezko, F.; Wrachtrup, J. Single-molecule scale magnetic resonance spectroscopy using quantum diamond sensors. *Rev. Mod. Phys.* **2024**, *96*, 025001.

8. Barry, J.F.; Schloss, J.M.; Bauch, E.; Turner, M.J.; Hart, C.A.; Pham, L.M.; Walsworth, R.L. Sensitivity optimization for NV-diamond magnetometry. *Rev. Mod. Phys.* **2020**, *92*, 015004.
9. Chen, X.D.; Ling, C.C.; Fung, S.; Beling, C.D.; Gong, M.; Henkel, T.; Tanoue, H.; Kobayashi, N. Beryllium implantation induced deep level defects in p-type 6H–silicon carbide. *J. Appl. Phys.* **2003**, *93*, 3117–3119.
10. van Duijn-Arnold, A.; Schmidt, J.; Poluektov, O.G.; Baranov, P.G.; Mokhov, E.N. Electronic structure of the Be acceptor centers in 6H-SiC. *Phys. Rev. B* **1999**, *60*, 15799.
11. Greulich-Weber, S.; Feege, M.; Kalinowski, J.; Gerstmann, U.; Rauls, E.; Overhof, H.; Spaeth, J.-M. Electron spin relaxation and coherence of nitrogen donors in silicon carbide. *Phys. Rev. B* **2003**, *67*, 195207.
12. Ramungul, N.; Chow, T.P. Current-controlled negative resistance (CCNR) in SiC PiN rectifiers. *IEEE Trans. Electron Devices* **1999**, *46*, 493–496.
13. Yaguchi, S.; Kimoto, T.; Ohyama, N.; Matsunami, H. Nitrogen ion implantation into 6H-SiC and application to high-temperature, radiation-hard diodes. *Jpn. J. Appl. Phys.* **1995**, *34*, 3036.
14. Banerjee, N.; Bell, C.; Ciccarelli, C.; Hesjedal, T.; Johnson, F.; Kurebayashi, H.; Moore, T.A.; Moutafis, C.; Stern, H.L.; Vera-Marun, I.J.; et al. Materials for quantum technologies: A roadmap for spin and topology. *Appl. Phys. Rev.* **2025**, *12*, 041328.
15. Awschalom, D.D.; Hanson, R.; Wrachtrup, J.; Zhou, B.B. Quantum technologies with optically interfaced solid-state spins. *Nat. Photonics* **2018**, *12*, 516–527.
16. Sekiguchi, Y. Quantum network technologies based on diamond NV centers. *JSAP Rev.* **2025**, *25*, 250420.
17. Sohr, P.; Koller, P.; Ecker, S.; Fink, M.; Scheidl, T.; Ursin, R.; Arshad, M.J.; Bonato, C.; Cilibrizzi, P.; Gali, A.; et al. Quantum communication networks with defects in silicon carbide. *arXiv* **2024**, arXiv:2403.03284.
18. Hu, G.; Huang, W.W.; Cai, R.; Wang, L.; Yang, C.H.; Cao, G.; Xue, X.; Huang, P.; He, Y. Single-electron spin qubits in silicon for quantum computing. *Intell. Comput.* **2025**, *4*, 0115.
19. Pogorzelski, J.; Horsthemke, L.; Homrighausen, J.; Stiegekötter, D.; Gregor, M.; Glösekötter, P. Compact and fully integrated LED quantum sensor based on NV centers in diamond. *Sensors* **2024**, *24*, 743.
20. Jiang, Z.; Cai, H.; Cernansky, R.; Liu, X.; Gao, W. Quantum sensing of radio-frequency signal with NV centers in SiC. *Sci. Adv.* **2023**, *9*, eadg2080.
21. Wolfowicz, G.; Heremans, F.J.; Anderson, C.P.; Kanai, S.; Seo, H.; Gali, A.; Galli, G.; Awschalom, D.D. Quantum guidelines for solid-state spin defects. *Nat. Rev. Mater.* **2021**, *6*, 906–925.
22. Morton, J.J.; Lovett, B.W. Hybrid solid-state qubits: The powerful role of electron spins. *Annu. Rev. Condens. Matter Phys.* **2011**, *2*, 189–212.
23. Onizhuk, M.; Galli, G. Colloquium: Decoherence of solid-state spin qubits: A computational perspective. *Rev. Mod. Phys.* **2025**, *97*, 021001.
24. Li, Y.; Li, H.; Yi, T.; Li, C.; Wei, J. A review of the study of diamond NV color centers: Fabrication, application and challenge. *Funct. Diamond* **2025**, *5*, 2567286.
25. Pacchioni, G. Spin qubits: Useful defects in silicon carbide. *Nat. Rev. Mater.* **2017**, *2*, 17052.
26. Jiang, S.; Li, Y.; Chen, Z.; Zhu, W.; Wu, Q.; He, H.; Wang, X. The effects of defects on the defect formation energy, electronic band structure, and electron mobility in 4H–SiC. *AIP Adv.* **2022**, *12*, 065328.
27. Frontiers in Physics Editorial Office. Fabrication and quantum sensing of spin defects in silicon carbide—A review. *Front. Phys.* **2023**, *11*, 1270602.
28. Ayedh, H.M.; Bobal, V.; Nipoti, R.; Hallén, A.; Svensson, B.G. Formation of carbon vacancy in 4H silicon carbide during high-temperature processing. *J. Appl. Phys.* **2014**, *115*, 013506.
29. Son, N.T.; Ivanov, I.G. Charge state control of the silicon vacancy and divacancy in silicon carbide. *J. Appl. Phys.* **2021**, *129*, 210903.
30. Ivády, V.; Davidsson, J.; Son, N.T.; Ohshima, T.; Abrikosov, I.A.; Gali, A. Identification of Si-vacancy related room-temperature qubits in 4H silicon carbide. *Phys. Rev. B* **2017**, *96*, 161114.
31. Singam, M.R.; Nesladek, M. Nanotesla Magnetometry with the Silicon Vacancy in Silicon Carbide. *Phys. Rev. Applied* **2021**, *15*, 064022.
32. Murzakhanov, F.F.; Sadovnikova, M.A.; Mamin, G.V.; Nagalyuk, S.S.; von Bardeleben, H.J.; Schmidt, W.G.; Biktagirov, T.; Gerstmann, U.; Soltamov, V.A. 14N hyperfine and nuclear interactions of axial and basal NV centers in 4H-SiC: A high frequency (94 GHz) ENDOR study. *J. Appl. Phys.* **2023**, *134*, 123906.
33. Latypova, L.R.; Gracheva, I.N.; Shurtakova, D.V.; Murzakhanov, F.F.; Sadovnikova, M.A.; Mamin, G.V.; Gafurov, M.R. Electron–nuclear interactions of NV defects in an isotopically purified 6H-28SiC crystal. *J. Phys. Chem. C* **2024**, *128*, 18559–18565.

34. Son, N.T.; Anderson, C.P.; Bourassa, A.; Awschalom, D.D.; Ivanov, I.G. Point defects in silicon carbide for quantum technologies. *Appl. Phys. Lett.* **2020**, *116*, 190501.
35. Abderrazak, H.; Hmida, E.S.B.H. Silicon carbide: Synthesis and properties. In *Properties and Applications of Silicon Carbide*; Gerlach, R., Ed.; IntechOpen: London, UK, 2011; pp. 361–388.
36. Evwaraye, A.O.; Smith, S.R.; Mitchel, W.C. Characterization of defects in n-type 6H-SiC single crystals by optical admittance spectroscopy. *MRS Online Proc. Libr.* **1994**, *339*, 711.
37. Holiatkina, M.; Solodovnyk, A.; Laguta, O.; Neugebauer, P.; Kalabukhova, E.; Savchenko, D. Nature of electrically detected magnetic resonance in highly nitrogen-doped 6 H-SiC single crystals. *Phys. Rev. B* **2024**, *110*(12), 125205. ~~Goldfarb, D.; Stoll, S. (Eds.) *EPR Spectroscopy: Fundamentals and Methods*; John Wiley & Sons: Chichester, UK, 2018~~.
38. Son, N.T.; Janzén, E.; Isoya, J.; Yamasaki, S. Hyperfine interaction of the nitrogen donor in 4H-SiC. *Phys. Rev. B* **2004**, *70*, 193207.
39. van Duijn-Arnold, A.; Zondervan, R.; Schmidt, J.; Baranov, P.G.; Mokhov, E.N. Electronic structure of the N donor center in 4H-SiC and 6H-SiC. *Phys. Rev. B* **2001**, *64*, 085206.
40. Murzakhanov, F.F.; Sadovnikova, M.A.; Mamin, G.V.; Shurtakova, D.Y.V.; Mokhov, E.N.; Kazarova, O.P.; Gafurov, M.R. Optical spin initialization of nitrogen vacancy centers in a 28Si-enriched 6H-SiC crystal for quantum technologies. *JETP Lett.* **2024**, *119*, 593–598.
41. Brill, T.B. Electric field gradient models based on the geometry of the coordination sphere. *J. Mol. Struct.* **1983**, *111*, 269–280.
42. Fabbro, G.; Pototschnig, J.; Saue, T. Beyond the Dailey–Townes model: Chemical information from the electric field gradient. *J. Phys. Chem. A* **2025**, *129*, 1006–1025.
43. Stoll, S.; Goldfarb, D. EPR interactions-nuclear quadrupole couplings. In *EPR Spectroscopy: Fundamentals and Methods*; Goldfarb, D., Stoll, S., Eds.; John Wiley & Sons: Chichester, UK, 2018; p. 95.
44. Xia, J.Y.; Liu, Z.T.; Liu, Q.J. First-principles investigation of 6H-SiC dominated by strong covalent bonding: Electronic structure, mechanical properties and optical properties. *J. Mol. Model*. **2025**, *31*, 292.
45. Smirnova, T.I.; Smirnov, A.I.; Clarkson, R.B.; Belford, R.L.; Kotake, Y.; Janzen, E.G. High-frequency (95 GHz) EPR spectroscopy to characterize spin adducts. *J. Phys. Chem. B* **1997**, *101*, 3877–3885.
46. Zvanut, M.E.; Van Tol, J. Nitrogen-related point defect in 4H and 6H SiC. *Physica B* **2007**, *401*, 73–76.
47. Chen, X.D.; Fung, S.; Beling, C.D.; Gong, M.; Henkel, T.; Tanoue, H.; Kobayashi, N. Beryllium implantation induced deep levels in 6H-silicon carbide. *Physica B* **2001**, *308*, 718–721.
48. Henkel, T.; Tanaka, Y.; Kobayashi, N.; Tanoue, H.; Hishita, S. Diffusion of implanted beryllium in silicon carbide studied by secondary ion mass spectrometry. *Appl. Phys. Lett*. **2001**, *78*, 231–233.
49. Lee, S.W.; Vlaskina, S.I.; Vlaskin, V.I.; Zaharchenko, I.V.; Gubanov, V.A.; Mishinova, G.N.; Podlasov, S.A. Silicon carbide defects and luminescence centers in current heated 6H-SiC. Semicond. *Phys. Quantum Electron. Optoelectron*. **2010**, *13*, 24–29.
50. Hawthorne, F.C.; Huminicki, D.M. The crystal chemistry of beryllium. *Rev. Mineral. Geochem.* **2002**, *50*, 333–403.
51. Shannon, R.D. Revised effective ionic radii and systematic studies of interatomic distances in halides and chalcogenides. *Acta Crystallogr. A* **1976**, *32*, 751–767.